\documentstyle[psfig]{l-aa}     % LaTeX A&A  Standard Fonts

\begin{document}
%+++++++++++++++++++++++++++++Abbreviations+++++++++++++++++++++++++++++++++++++
%
%   Masses and luminsosities
%
\newcommand{\mo}    {\hbox{${\rm M}_{\odot}$}}
\newcommand{\mhtwo} {\hbox{$M_{{\rm H}_2}$}}
\newcommand{\mhi}   {\hbox{$M_{\rm HI}$}}
\newcommand{\mdust} {\hbox{$M_{\rm d}$}}
\newcommand{\mdyn}  {\hbox{$M_{\rm dyn}$}}
\newcommand{\lo}    {\hbox{${\rm L}_{\odot}$}}
\newcommand{\lb}    {\hbox{$L_{\rm B}$}}
\newcommand{\lbo}   {\hbox{$L_{\rm BOL}$}}
\newcommand{\lfir}  {\hbox{$L_{\rm FIR}$}}
\newcommand{\lco}   {\hbox{$L_{\rm CO}$}}
\newcommand{\halfa} {\hbox{$H\alpha$}}
\newcommand{\hbeta} {\hbox{$H\beta$}}
\newcommand{\lhalfa}{\hbox{$L_{\rm H\alpha}$}}
\newcommand{\jykms} {\hbox{Jy\,km\,s$^{-1}$}}
\newcommand{\sdoz}  {\hbox{$S_{12\mu \rm m}$}}           % s(12)
\newcommand{\stwe}  {\hbox{$S_{25\mu \rm m}$}}           % s(25)
\newcommand{\ssix}  {\hbox{$S_{60\mu \rm m}$}}           % s(60)
\newcommand{\shun}  {\hbox{$S_{100\mu \rm m}$}}          % s(100)
\newcommand{\micron}{$\mu\hbox{m}$}
%
%   Column densities
%
\newcommand{\nhi}   {\hbox{$N_{\rm HI}$}}                % NHI
\newcommand{\Nhtwo} {\hbox{$N_{{\rm H}_2}$}}             % NH2
%
%   Velocities and angles
%
\newcommand{\kms}   {\hbox{${\rm km\,s}^{-1}$}}
\newcommand{\ms}    {\hbox{${\rm m\,s}^{-1}$}}
\newcommand{\kmsmpc}{\hbox{${\rm km\,s}^{-1}{\rm Mpc}^{-1}$}}
\newcommand{\vlsr}  {\hbox{$v_{\rm LSR}$}}
\newcommand{\vhel}  {\hbox{$v_\odot$}}
\newcommand{\delv}  {\hbox{$\Delta v_{1/2}$}}
\newcommand{\dv}    {\hbox{$\Delta v$}}
\newcommand{\kmspc} {\hbox{km\,s$^{-1}$\,pc$^{-1}$}}
\newcommand{\asec}  {\hbox{$^{\prime\prime}$}}
\newcommand{\amin}  {\hbox{$^{\prime}$}}
%
%   Temperatures
%
\newcommand{\tex}   {\hbox{$T_{\rm x}$}}
\newcommand{\tmb}   {\hbox{$T_{\rm mb}$}}
\newcommand{\tb}    {\hbox{$T_{\rm B}$}}
\newcommand{\tkin}   {\hbox{$T_{\rm k}$}}
\newcommand{\trec}  {\hbox{$T_{\rm rec}$}}
\newcommand{\tsys}  {\hbox{$T_{\rm sys}$}}
\newcommand{\tdust} {\hbox{$T_{\rm d}$}}
\newcommand{\tastar}{\hbox{$T_{\rm A}^{*}$}}
%
%   Units and constants
%
\newcommand{\hub}   {\hbox{$H_{\rm 0}$}}
\newcommand{\pac}   {\hbox{${\rm pc}$}}
\newcommand{\kpc}   {\hbox{$\rm kpc$}}
\newcommand{\mpc}   {\hbox{$\rm Mpc$}}
\newcommand{\ico}   {\hbox{$I_{\rm CO}$}}
\newcommand{\nhtwo} {\hbox{\hbox{$n({\rm H}_2)$}}}
\newcommand{\mocpc} {\hbox{${\rm M}_\odot\,\hbox{\rm pc}^{-3}$}}
\newcommand{\mopcsq}{\hbox{${\rm M}_\odot\,\hbox{\rm pc}^{-2}$}}
\newcommand{\sfrat} {\hbox{${\rm M}_\odot\,\hbox{\rm yr}^{-1}$}}
\newcommand{\kkms}  {\hbox{${\rm K\,km\,s}^{-1}$}}
\newcommand{\cm}    {\hbox{${\rm cm}$}}
\newcommand{\cmsq}  {\hbox{${\rm cm}^{-2}$}}
\newcommand{\cmcb}  {\hbox{${\rm cm}^{-3}$}}
\newcommand{\hI}    {\hbox{${\rm HI}$}}
\newcommand{\hII}   {\hbox{${\rm HII}$}}
%
% molecules
%
\newcommand{\htwo}  {\hbox{${\rm H}_2$}}                % H2
\newcommand{\twco}  {\hbox{${\rm ^{12}CO}$}}            % 12CO
\newcommand{\thco}  {\hbox{$^{13}{\rm CO}$}}            % 13CO
\newcommand{\htwoco}{\hbox{${\rm H}_2{\rm CO}$}}        % H2CO
\newcommand{\jone}  {$J$=1$-$0}                         % J=1-0
\newcommand{\jtwo}  {$J$=2$-$1}                         % J=2-1
\newcommand{\jthree}{$J$=3$-$2}                         % J=3-2
\newcommand{\jeno}  {$J$=1$\leftarrow$0}                % J=0-1
\newcommand{\jowt}  {$J$=2$\leftarrow$1}                % J=1-2
\newcommand{\jeerht}{$J$=3$\leftarrow$2}                % J=2-3
\newcommand{\jhtruof}{$J$=4$\leftarrow$3}               % J=3-4
\newcommand{\jevif}{$J$=5$\leftarrow$4}                 % J=4-5
%
%  Greeks
%
\newcommand{\alfa}{$\alpha $}                           % gr. alpha
\newcommand{\bet} {$\beta $}                            % gr. beta
\newcommand{\gam} {$\gamma $}                           % gr. gamma
\newcommand{\del} {$\delta $}                           % gr. delta
\newcommand{\eps} {$\epsilon $}                         % gr. epsilon
\newcommand{\my}  {$\mu$}                               % gr. my
%\newcommand{\tau}{$\tau$}                              % gr. tau
%
%	Various useful things
%
\newcommand{\alde}{($\Delta \alpha ,\Delta \delta $)}
\newcommand{\ffam}{\hbox{$\,.\!\!^{\prime}$}}
\newcommand{\ffas}{\hbox{$\,.\!\!^{\prime\prime}$}}
\newcommand{\ffM} {\hbox{$\,.\!\!^{\rm M}$}}
\newcommand{\ffm} {\hbox{$\,.\!\!^{\rm m}$}}
\newcommand{\fft} {\hbox{$\,.\!\!\!^{M}$}}
\newcommand{\ffd}{\hbox{$\,.\!\!^{\circ}$}}
%++++++++++++++++++++++++++++++++++++++++++++++++++++++++++++++++++++++++++++++

\topmargin=0.0cm
\thesaurus{ 09.13.2;  % ISM                  : molecules
            11.05.1;  % Galaxies             : elliptical and lenticular, cD
            11.09.1;  % Galaxies             : indivdual: NGC759
            11.09.4;  % Galaxies             : ISM
            11.11.1;  % Galaxies             : kinematics and dynamics
	         }
%%%%%%%%%%%%%%%%%%%%%%%%%%%%%%%%%%%%%%%%%%%%%%%%%%%%%%%%%%%%%%%%%%%%%%%%%%%%%%%%
   \title{Molecular gas in the elliptical galaxy NGC\,759}

   \subtitle{Interferometric CO observations}
%%%%%%%%%%%%%%%%%%%%%%%%%%%%%%%%%%%%%%%%%%%%%%%%%%%%%%%%%%%%%%%%%%%%%%%%%%%%%%%%
%
   \author{T.~Wiklind \inst{1},  F.~Combes \inst{2}, C.~Henkel \inst{3},
            F.~Wyrowski \inst{3,4}}
   \offprints{T.~Wiklind, tommy@oso.chalmers.se}
   \institute{Onsala Space Observatory,
              S--43992 Onsala, Sweden
   \and
              DEMIRM
              Observatoire de Paris, 61 Avenue de l'Observatoire,
              F--75014 Paris, France
   \and
              Max-Planck-Institut f\"{u}r Radioastronomie,
              Auf dem H\"{u}gel 69, D--53121 Bonn, Germany
   \and
              Physikalisches Institut der Universit\"{a}t zu K\"{o}ln,
              Z\"{u}lpicher Strasse 77, D--50937 K\"{o}ln, Germany
             }
   \date{Received date; Accepted date}
   \maketitle
%%%%%%%%%%%%%%%%%%%%%%%%%%%%%%%%%%%%%%%%%%%%%%%%%%%%%%%%%%%%%%%%%%%%%%%%%%%%%%%%
%
\begin{abstract}
We present interferometric observations of CO(1--0) emission in
the elliptical galaxy NGC\,759 with an angular resolution of
3\ffas1$\times$2\ffas3 ($990 \times 735$\,pc at a distance of
66\,Mpc).
NGC\,759 contains $2.4 \times 10^{9}$\,\mo\ of molecular gas.
Most of the gas is confined to a small circumnuclear ring with a
radius of 650\,pc with an inclination of $40^{\circ}$.
The maximum gas surface density in the ring is
750\,\mopcsq. Although this value is very high, it is always
less than or comparable to the critical gas surface density for
large scale gravitational instabilities.
The CO \jtwo/\jone\ line ratio is low (0.4), implying sub-thermal
excitation.
This low ratio is consistent with a two--component molecular gas,
consisting of a cold and dense phase, containing most of the mass,
and a warm and diffuse gas component, dominating the CO emission.
Compared to galaxies of similar gas surface densities,
NGC\,759 is underluminous in \lfir\ with respect to its molecular
gas mass, suggesting that the star forming efficiency in NGC\,759
is low.
We discuss the possibility that the molecular gas and current
star formation activity in NGC\,759 could be signatures of a
late stage of a merging between two gas--rich disk galaxies.
We use a mass model of the underlying galaxy which is applicable
to spherical galaxies with an $r^{1/4}$--luminosity profile when
interpreting our CO data. This leads to more modest estimates of
the molecular gas mass fraction and the surface gas density than
would have been derived using simpler models, suggesting that
many of the spectacular molecular gas properties of ultraluminous
FIR galaxies, which could be described by similar mass distributions,
may have to be revised.
 
\keywords{ISM: molecules -- Galaxies: elliptical and lenticular, cD
          -- Galaxies: individual: NGC759 -- Galaxies: ISM --
          Galaxies: kinematics and dynamics}

\end{abstract}
 
%________________________________________________________________ 
\section{Introduction}

Faber \& Gallagher (1976) found the apparent absence of an interstellar
medium (ISM) in elliptical galaxies surprising since mass loss from
evolved stars should contribute as much as $10^{9}-10^{10}$\,\mo\ of
gas to the ISM during a Hubble time.  Today we know that ellipticals
do contain an ISM, albeit very different from that found in spiral
galaxies. The most massive ISM component is a hot X--ray radiating halo,
with a typical mass of $10^{8}-10^{10}$\,\mo\ (Fabbiano 1990). Small
amounts of photoionized gas (typically $10^{2}-10^{4}$\,\mo) can be found
in about 60\% of all ellipticals (Caldwell 1984, Phillips et al. 1986).
In addition, $\sim45$\% of a magnitude limited sample of ellipticals
were detected at both 60 and 100\,\micron\ by IRAS (Jura et al. 1987,
Knapp et al. 1989), suggesting the presence of a significant dust component.
Atomic hydrogen gas is seen in $\sim15$\% of all ellipticals (Knapp et al. 
1985). A similar detection rate has been found for molecular gas, seen 
through CO emission (Lees et al. 1991, Wiklind et al. 1995).

Elliptical galaxies have less molecular gas per blue luminosity than
spirals and a much larger dispersion in their \lb/\mhtwo\ ratios.
Wiklind et al. (1995) found that the molecular gas in ellipticals
is {\em not} correlated with the old stellar population and, hence,
is accreted from an external source. A similar result for the atomic
hydrogen gas was reached by Knapp et al. (1985).

The discovery of kinematically decoupled cores in elliptical galaxies
has strongly favoured the idea that these systems can be formed
through the merging of two disk galaxies of similar sizes (Franx
\& Illingworth 1988,  Rix \& White 1992). The main stellar component
is hot and slowly rotating while a small centrally
concentrated and kinematically distinct core exhibits rapid rotation. 
Numerical simulations of mergers between two disk galaxies successfully 
reproduce the observed properties (Hernquist \& Barnes 1991).
The dissipative gas is concentrated in the center of the resulting
galaxy and  the kinematically decoupled core is observable in both
ionized and molecular gas (cf. Bertola \& Bettoni 1988, Balcells
\& Stanford 1990, Wang et al. 1992). The early stages of mergers
are accompanied by intense and efficient star  formation, giving a
chaotic appearance to the system. The underlying old stellar
population, however, has a luminosity profile which in most
cases closely  follows the de Vaucouleurs $r^{1/4}$ law.
Hence, once the gas has been consumed and star formation ceases,
the resulting system will not only resemble but actually be an
elliptical galaxy. 

One elliptical with relatively strong CO emission is NGC\,759
(Wiklind et al. 1995). We have chosen this galaxy for high resolution
observations with the aim of resolving the distribution and
kinematical structure of the molecular gas. In this paper we
present our results and discuss the rather peculiar conditions
under which the molecular gas exists in this galaxy. 

%1
\begin{figure*}
\psfig{figure=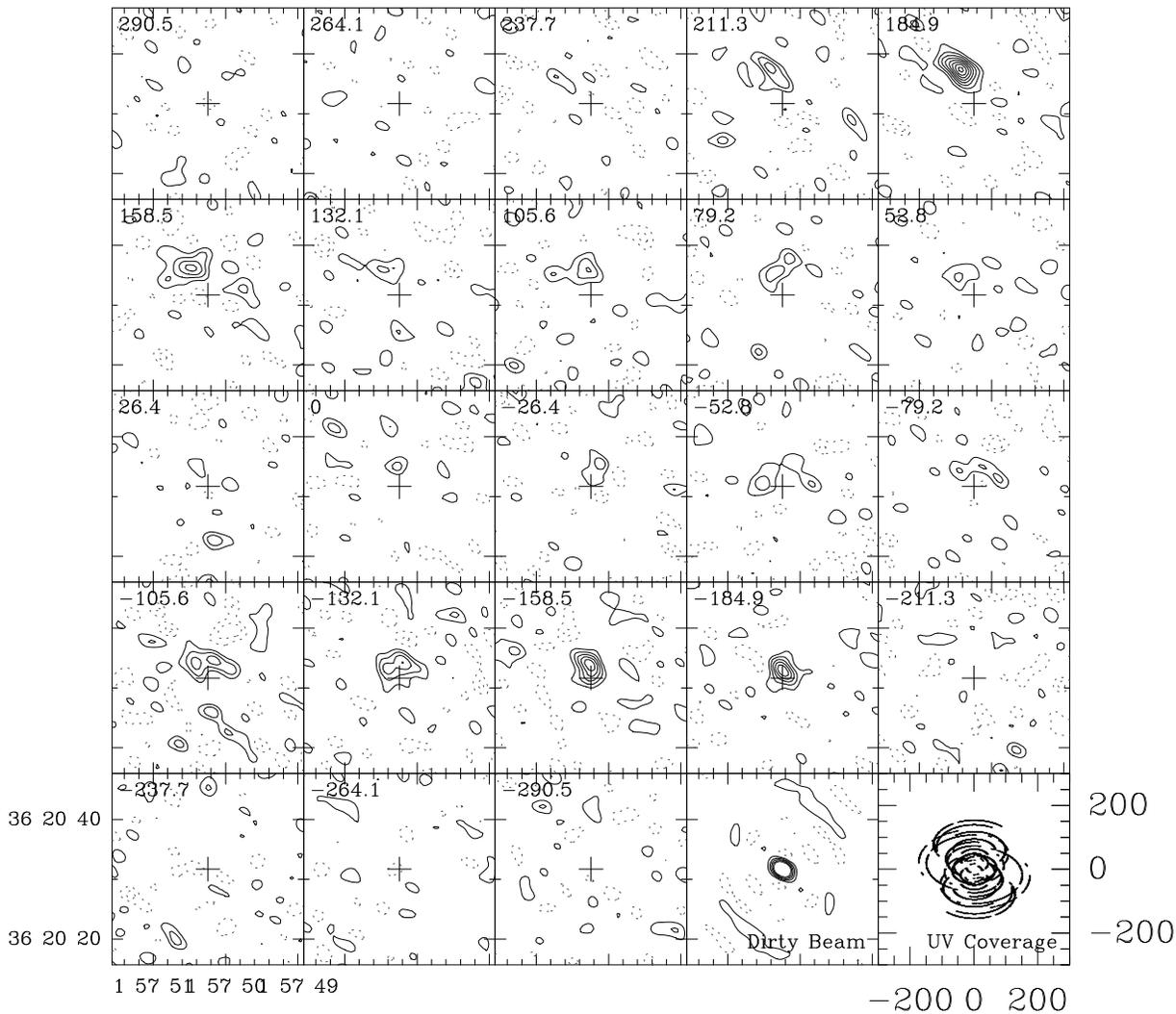,bbllx=15mm,bblly=105mm,bburx=190mm,bbury=272mm,width=15.5cm,angle=0}
%\picplace{14.0cm}
\caption[]{Channel maps of the CO emission in NGC\,759, separated by
10\,MHz (26.4\,\kms). Velocity offsets are relative to the systemic
velocity of 4665\,\kms. The cross in each map corresponds
to the phase reference center.
The contour levels are in steps of 10\,mJy (0.13\,K). Also shown are the dirty
beam and the uv--coverage. The latter is in units of meters. Coordinates are
J2000.0.}
\end{figure*}

%2
\begin{figure*}
\psfig{figure=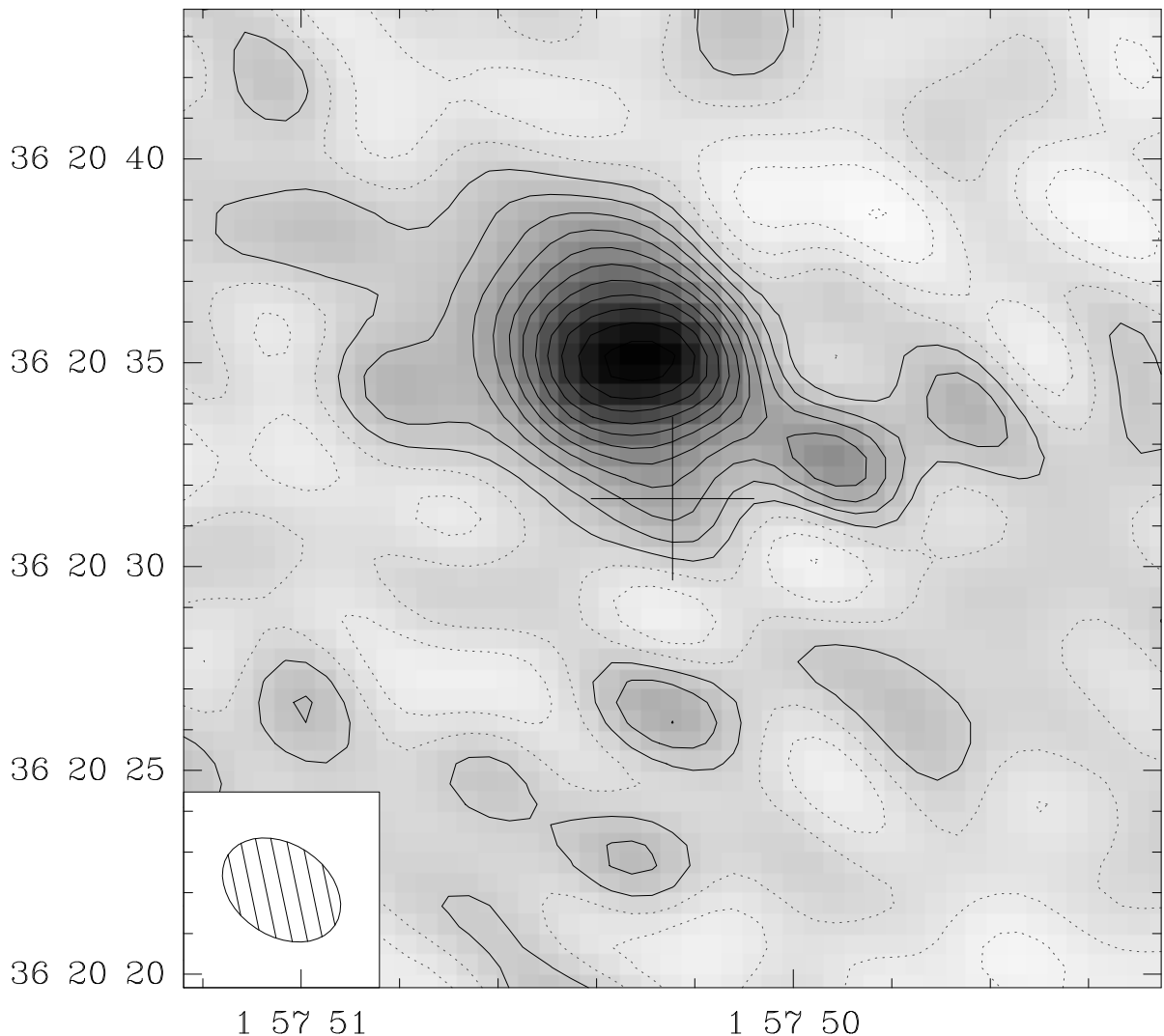,bbllx=10mm,bblly=18mm,bburx=185mm,bbury=190mm,width=15.5cm,angle=0}
%\picplace{10.0cm}
\caption[]{Total integrated CO(1--0) emission obtained by adding all channels
between $-237$\,\kms\ and $+185$\,\kms\ (relative to the systemic velocity).
The level spacing is 0.8\,\jykms\ and the rms noise in the map is 0.8\,\jykms.
The cross marks the phase reference center and the optical position, but
is offset a few arcsec from the CO and radio continuum peaks (see text for
details). Coordinates are J2000.0.}
\end{figure*}

%--------------------------Table 1--------------------------------------------
\begin{table}
\begin{flushleft}
\caption[]{Data on NGC\,759}
\scriptsize
\begin{tabular}{|r|cc|c|}
\hline
 & & & \\
Type                & E0 (E1)                &           & (1) \\
Size                & 1\ffam6$\times$1\ffam4 &           & (1) \\
R.A. (2000)         & \ $01^{\rm h}57^{\rm m}50.3^{\rm s}$ & & (2) \\
DEC (2000)          & $+36^{\circ}20\amin35\asec$ &      & (2) \\
V$_{\odot}$         & 4665                   & \kms      & (3,4) \\
Distance$^{a)}$     & 66                     & Mpc       & (3) \\
$R_{25}$            & 14.5                   & kpc       & (4) \\
$B_{\rm T}^{0}$     & 13.24                  &           & (1) \\
$M_{\rm B}$         & $-$20.86               &           & (3) \\
$(B-V)_{\rm T}^{0}$ & 1.00                   &           & (1) \\
$(U-B)_{\rm T}^{0}$ & 0.52                   &           & (1) \\
S$_{12}$            & $<102$                 & mJy       & (5) \\
S$_{25}$            & $<132$                 & mJy       & (5) \\
S$_{60}$            & $732\,\pm66$           & mJy       & (5) \\
S$_{100}$           & $2076\,\pm187$         & mJy       & (5) \\
S$_{1.4GHz}$        & $16.4$                 & mJy       & (6) \\
 & & & \\
\hline
 & & & \\
\lb                 & $3.46 \times 10^{10}$  & \lo       & (4) \\
\lfir               & $1.13 \times 10^{10}$  & \lo       & (4) \\
\lco                & $5.03 \times 10^{8}$   & \kkms\,kpc$^{-2}$ & (4) \\
 & & & \\
\hline
 & & & \\
\mhtwo              & $ 2.4 \times 10^{9}$   & \mo       & (4) \\
\mhi                & $<7.4 \times 10^{8}$   & \mo       & (7) \\
\mdust              & $ 3.5 \times 10^{6}$   & \mo       & (4) \\
 & & & \\
\hline
 & & & \\
\tdust$^{b)}$       & $31$                   & K         & (4) \\
SFR$^{c)}$          & $7$                    & \sfrat    & (3) \\
\lfir/\mhtwo        & $4.7$                  & \lo/\mo   & (4) \\
\lfir/\lb           & $0.3$                  & \lo/\lbo  & (4) \\
 & & & \\
\hline
\end{tabular}
\\
(1) \ RC3 (de Vaucouleurs et al. 1991). \\
(2) \ Peak of the CO and radio continuum emission (cf. Fig.\,2). \\
(3) \ Adopted in this paper. \\
(4) \ Wiklind et al. (1995). \\
(5) \ IRAS Faint Source Catalog. \\
(6) \ Feretti \& Giovannini (1994). \\
(7) \ Huchtmeier et al. (1995). \\
\ \\
a)\ Derived from the mean heliocentric velocity of the A262 cluster
$4845 \pm 438$ (Moss \& Dickens 1977) and a Hubble constant of 75\,\kmsmpc,
corrected for the solar motion relative to the Local Group and the peculiar
motion of the Local Group with respect to the Virgo cluster (Aaronson et al. 
1982). \\
b)\ Derived using a $\nu^1$ emissivity law. The dust temperature is slightly lower
than in ref (4) due to modified FIR fluxes. \\
c)\ Derived using $SFR \approx 6.5 \times 10^{-10}$\,\lfir\ \sfrat.
\end{flushleft}
\end{table}
%----------------------End Table 1--------------------------------------------

%--------------------------Table 2--------------------------------------------
\begin{table*}
\begin{flushleft}
\caption[]{Observing parameters}
\small
\begin{tabular}{c|cr|cc}
\hline
 & & & & \\
\multicolumn{1}{c|}{Date}                     &
\multicolumn{1}{c}{Configuration$^{a)}$}      &
\multicolumn{1}{c|}{Hours}                    &
\multicolumn{1}{c}{Bandpass calibrator}       &
\multicolumn{1}{c}{Phase calibrator}          \\
 & & & & \\
\hline
 & & & & \\
   27 June 1993      & N05\ N13\ E10        &  8.5 & 3C345   & 0234+285       \\
  1--2 Jul 1993      & N05\ N13\ E10\ W12   &  9.2 & 3C454.3 & 0234+285       \\
19--20 Aug 1993      & E03\ W05\ N05        &  7.4 & 3C84    & 3C84           \\
    21 Aug 1993      & E03\ N05\ W00        &  5.0 & 3C84    & 3C84           \\
31 Aug -- 1 Sep 1993 & E10\ W09\ N15\ N03   & 10.7 & 3C84    & 3C84           \\
  3--4 Sep 1993      & E10\ W09\ N15\ N03   &  9.6 & 3C84    & 3C84           \\
    29 Mar 1994      & E03\ N05\ W00\ W05   &  2.2 & 3C345   & 0133+476, 3C84 \\
 & & & & \\
\hline
\end{tabular}
\\
a)\ N, W and E refer to the northern, western and eastern arm, respectively.
Numbers denote \\ the distance from the intersection of the arms (W00) in units
of 8\,m.
\end{flushleft}
\end{table*}
%----------------------End Table 2-------------------------------------------- 

%3
\begin{figure}
\psfig{figure=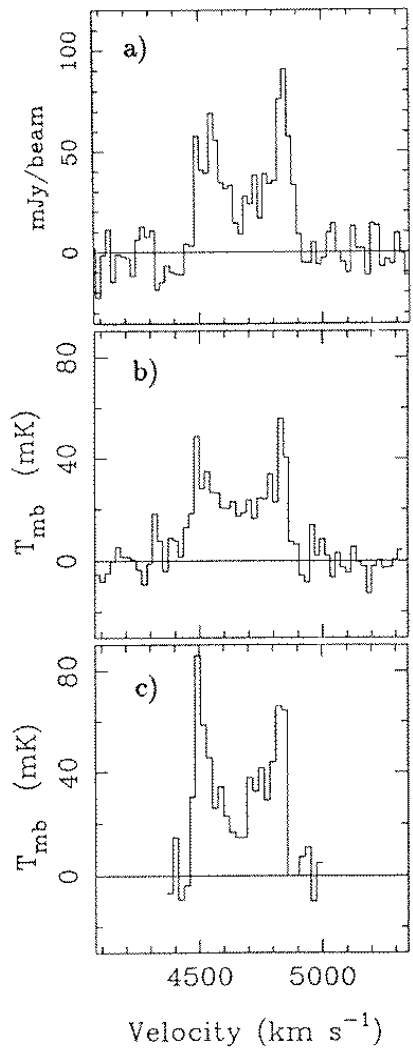,bbllx=80mm,bblly=65mm,bburx=150mm,bbury=225mm,width=6.5cm,angle=0}
%\picplace{14.0cm}
\caption[]{{\bf a)}\ CO(1--0) spectrum integrated over the
emission region seen in Fig.\,2.
{\bf b)}\ Single dish CO(1--0) spectrum obtained with with the
IRAM 30--m telescope. {\bf c)} Single dish CO(2--1) spectrum
obtained with the IRAM 30--m telescope.
}
\end{figure} 

%4
\begin{figure}
\psfig{figure=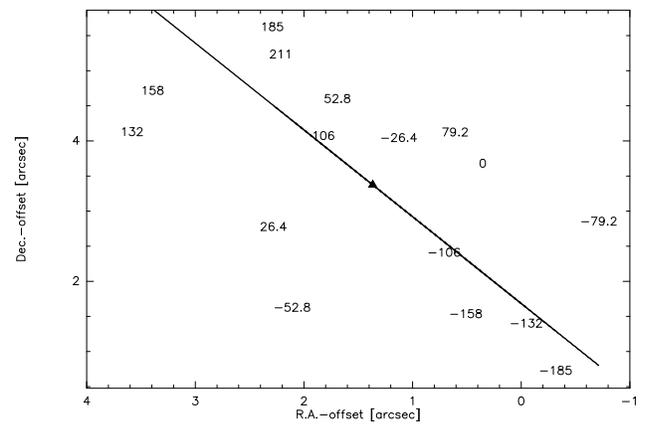,bbllx=15mm,bblly=10mm,bburx=200mm,bbury=275mm,width=8.5cm,angle=-90}
%\picplace{5.0cm}
\caption[]{Positions of Gaussian fits to the individual 10\,MHz
channel maps relative to the phase reference center. The triangle
represents the peak of the integrated CO emission seen in Fig.\,2
and coincides with the kinematical center.
The line is a best fit position angle to the kinematical data.}
\end{figure}

%5
\begin{figure}
\psfig{figure=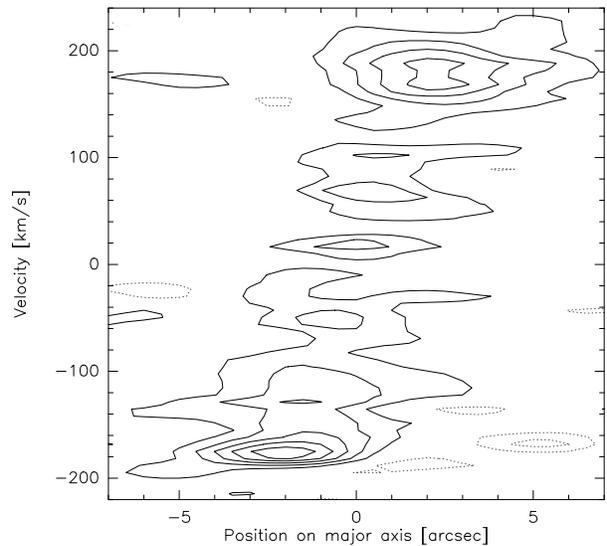,bbllx=20mm,bblly=55mm,bburx=200mm,bbury=250mm,width=8.3cm,angle=-90}
%\picplace{5.0cm}
\caption[]{Position--velocity diagram along the major axis indicated in
Fig.\,4. The level spacing is 8\,mJy/beam. The offsets are measured relative
to the position of maximum integrated emission (cf. Fig.\,2). Velocities
are relative to the systemic velocity.}
\end{figure}

%6
\begin{figure}
\psfig{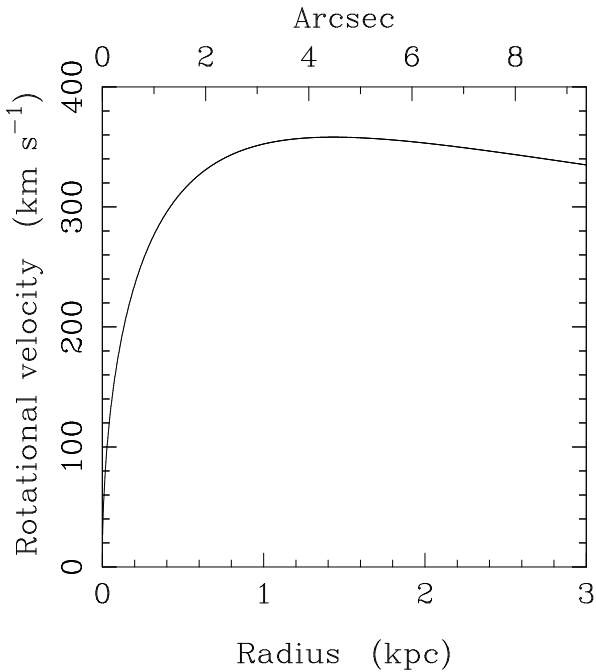}
%\picplace{5.0cm}
\caption[]{The rotation curve corresponding to a spherical galaxy with
an $r^{1/4}$ projected luminosity profile (Sect.\,5.1). The angular scale on top
corresponds to a distance of 66\,Mpc.}
\end{figure}

\section{Abell 262 and NGC\,759}

NGC\,759, a giant E0 galaxy, is one of the brightest members of
the cluster A262 and forms a noninteracting pair with NGC\,753.
The cluster A262 is comparable to the Virgo cluster in its richness
and has a similar X--ray luminosity.
The X--ray emission is centered on the elliptical NGC\,708, which appears to be
at rest within the cluster (Jones \& Forman 1984). NGC\,708 is a radio source
with a dust lane oriented almost perpendicular to its radio axis
(Ebneter \& Balick 1985). Its far--infrared (FIR) luminosity derived
from IRAS data is $3\times10^{9}$\,\lo. A deep search for CO emission
in NGC\,708 gave an upper  limit to the molecular gas mass of
$4\times10^{7}$\,\mo\ (Braine \& Dupraz 1994).

An optical spectroscopic survey of ten early--type galaxies
in A262 (including NGC\,708 and NGC\,759) showed that only NGC\,759 has a
spectrum showing recent star formation activity (Vigroux et al. 1989).
A recent study of NGC\,759 (E. Davoust private communication) shows
emission lines of \halfa, NII and SII. The FIR luminosity of NGC\,759 is
$1.1 \times 10^{10}$\,\lo\ and we observe a CO luminosity corresponding to
an \htwo\ mass of $2.4 \times 10^{9}$\,\mo\ (Wiklind et al. 1995; Table\,1).
Despite the presence of a massive cold ISM component, NGC\,759 is
a true giant elliptical: It has an $r^{1/4}$ light profile (i.e. Sandage
\& Perelmuter 1991) and it falls upon the fundamental plane relationship
for elliptical galaxies (c.f. Djorgovski \& Davis 1987; Guzm\'{a}n et al.
1993).

\section{Observations and data reduction}

\subsection{Observations}

Synthesis observations of NGC\,759 were made with the Institute de Radio
Astronomie Millim\`{e}trique (IRAM) interferometer at Plateau de Bure (PdB),
situated at an altitude of 2560\,m.
The PdB array comprises 4 telescopes of 15--m diameter which can be placed
at 26 different stations along a T--shaped runway extending 288\,m East--West
and 160\,m North--South (for a detailed description, see Guilloteau et al. 
1992). NGC\,759 was observed with the standard CD configuration (see Table\,2).
Maximum and minimum baselines were 160 and 24\,m, respectively.
The observations were made on 7 occasions between June 1993 and March 1994,
with a total observing time of 52.5 hours.
The SIS receivers were single--sideband tuned to the redshifted
$^{12}$CO, \jone\  frequency 113.508\,GHz, with the CO line placed in the upper
sideband. The receiver temperatures were 40--140\,K, resulting in system
temperatures of 300--800\,K (\tastar). The correlator was
configured as five slightly overlapping $64\times2.5$\,MHz subbands,
giving a velocity resolution of 6.6\,\kms\ and a total bandwidth of 1340\,\kms.

\subsection{Data reduction}

The CLIC programme and associated Grenoble software were used for all stages
of the data reduction. Bad scans were flagged, followed by bandpass 
calibration using strong continuum sources observed in the beginning of 
each observing run. Each correlator subband was fitted separately with 
a 7--10th order polynomial. The edges and two channels in the center of 
the subband were affected by the Gibbs phenomenon and were omitted in the 
fitting process. The errors in the amplitude and
phases for the bandpass correlations were $2-8$\% and $1.2-4^{\circ}$,
respectively.

Cubic splines were fitted to the measured phases of the phase calibrators.
In a few cases a focus measurement during the observations introduced a phase
jump. In order to match these discontinuities we applied two fits separately.
Due to bad weather conditions the observations in 
June 27, August 31, and September
3 had residual phase errors of $25-35^{\circ}$. The rest of the observations
have residual phase errors of $5-15^{\circ}$.

Long integrations on either 3C345, 3C454.3 or 3C84 were made at the beginning
of each observing run. These sources were used as bandpass calibrators as well
as flux references.
Since most calibrator sources at millimeter wavelengths are variable,
it is difficult to apply a correct flux scale. The flux values were
bootstrapped from measurements of sources with accurately known fluxes. 
For 3C84 we used $5.1 \pm 0.8$\,Jy and for 3C454.3 $6.9 \pm 0.8$\,Jy,
derived with the IRAM 30--m telescope in August 1993. We could not
get a reasonable solution using $6.2$\,Jy for 3C345, derived in February 1993.
Instead we used 0235+285, with $2.05$\,Jy derived in July 1993. For the March 29
1994 observation 0133+476 was used as a flux calibrator ($1.45$\,Jy).
The adopted fluxes were used to calculate the antenna gains.
With these gains the fluxes of all phase calibrators were determined, only
using scans with the highest amplitudes to ensure that no phase fluctuations
had  reduced the amplitude. Cubic splines were then fitted to the NGC\,759 data.

From the calibrated data we constructed {\em uv}--tables, applied a fast Fourier
transformation and deconvolved the resulting maps with the CLEAN algorithm
using natural weighting. The resulting size of the clean beam is
3\ffas1$\times$2\ffas3, corresponding to $990 \times 735$\,pc at a distance
of 66\,Mpc. The dirty beam and the {\em uv}--coverage are shown as part of Fig.\,1.

%7
\begin{figure}
\psfig{figure=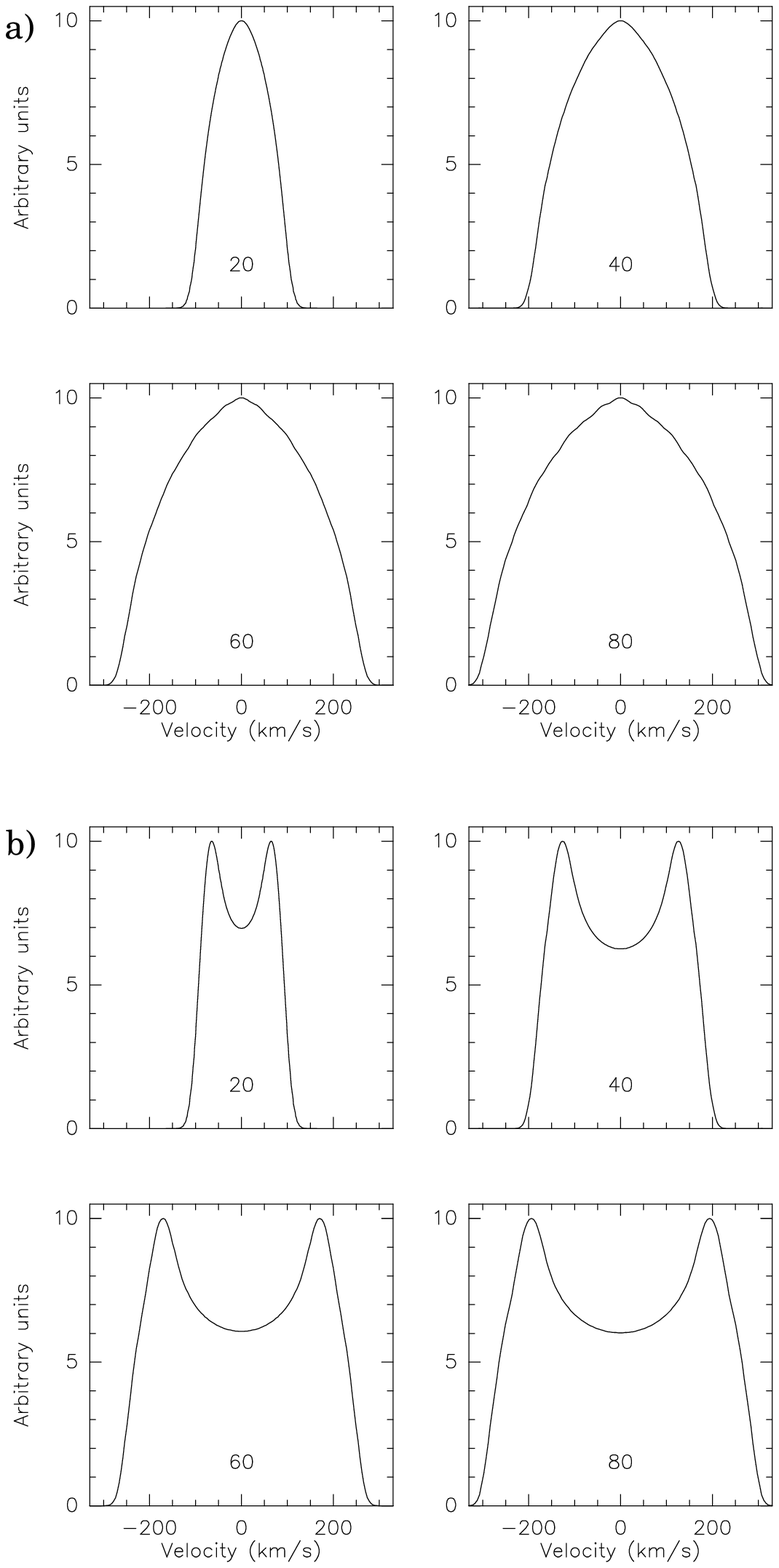,bbllx=45mm,bblly=40mm,bburx=170mm,bbury=270mm,width=8.0cm,angle=0}
%\picplace{6.0cm}
\caption[]{Synthetic spectra of an unresolved source, derived using the
formalism presented in Sect.\,5.3.
{\bf a)} A filled disk and, {\bf b)} a ring.
Only the ring is compatible with the observed emission profile from NGC\,759.
The numbers in the spectra correspond to the inclination of the gas distribution.}
\end{figure}

%8
\begin{figure}
\psfig{figure=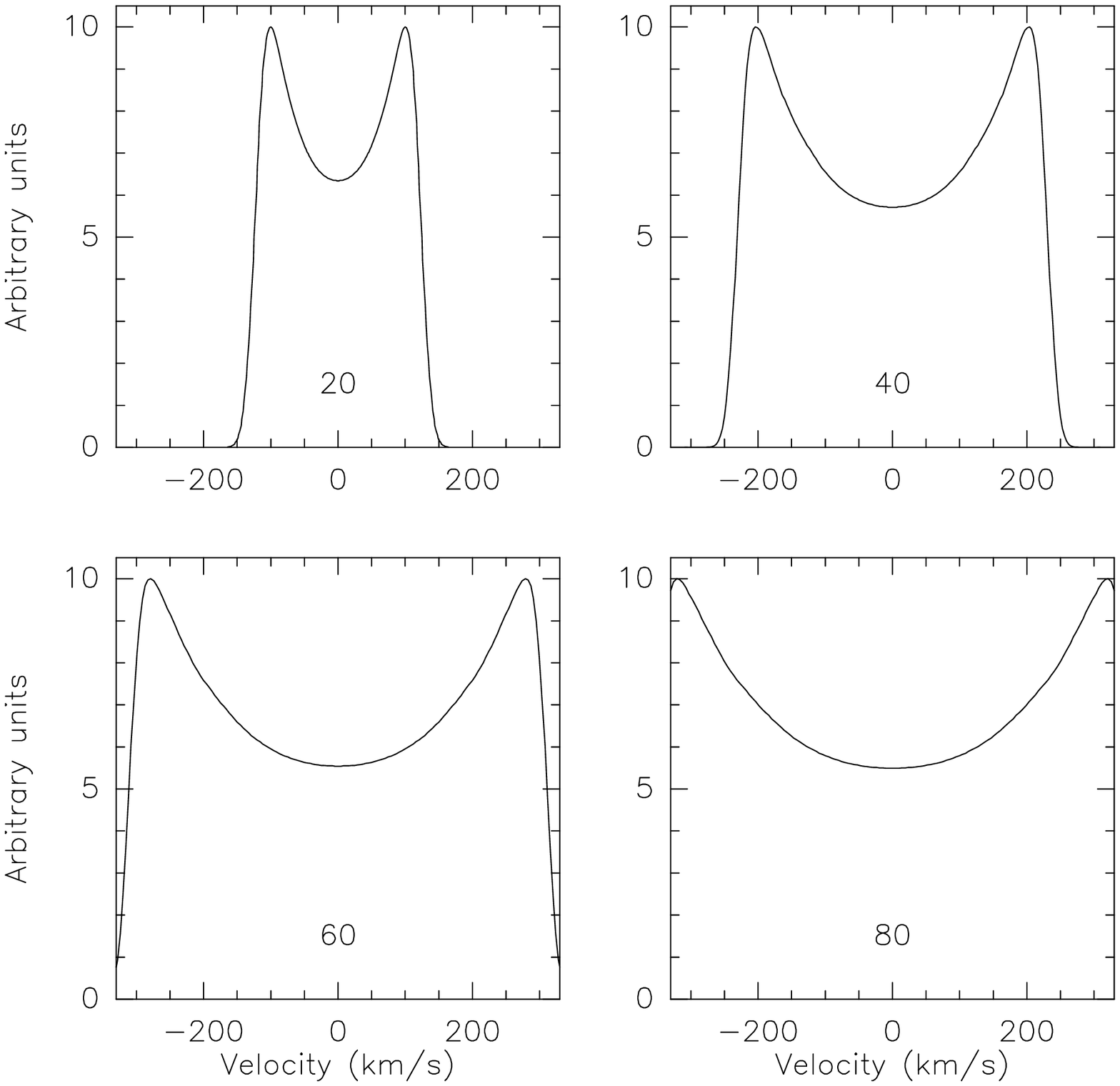,bbllx=10mm,bblly=5mm,bburx=205mm,bbury=195mm,width=8.0cm,angle=-90}
%\picplace{6.0cm}
\caption[]{Synthetic spectra obtained from the model gas distribution
of NGC\,759 (Sect.\,5.3), viewed at different inclinations.}
\end{figure}

\section{Results}

\subsection{The CO distribution}

Individual channel maps with a velocity separation of 26.4\,\kms\ (10\,MHz)
are shown in Fig.\,1. CO emission is clearly detected in channels approximately
$\pm 200$\,\kms\ around the systemic velocity of 4665\,\kms. The 
strongest emission
is seen around $+185$ and $-185$\,\kms\ relative to the systemic velocity, with peak
flux densities of 95 and 65\,mJy per beam. The rms noise in the
individual maps is 6\,mJy per beam.

A map of the total integrated CO emission was made by adding all the
channels between $-237$\,\kms and $+185$\,\kms. The result is presented
in Fig.\,2. The peak of the CO emission is 9.2\,\jykms. By summing the
same number of channels without emission we obtained an rms noise of
0.8\,\jykms.

Two things are immediately evident from Fig.\,2.
First of all, the peak of the CO emission is offset $\sim4$\asec\ from
the nominal pointing of the telescope. Since the peak of the CO
emission also corresponds to the kinematical center, as well as
the center of radio continuum emission (Feretti \& Giovannini 1994),
this offset most likely reflects an uncertainty in the optical
coordinates.
Secondly, the emission region has an elliptical shape, with an extent
of 5\ffas4$\times$4\ffas3 ($1.8\times1.4$\,kpc) at the 50\% level.
This is larger than the synthesized beam which has a FWHM size of
3\ffas1$\times$2\ffas3. However, the synthesized beam and the CO emission
region have similar ratios of the major and minor axis as well as the
same position angle ($\sim50^{\circ}$). This could be a coincidence or
an indication that our synthesized beam is larger than estimated.
In addition to the bright compact component, there is low level emission
extending towards the South--West, to a distance of $\sim7\asec$ from
the center.

At a frequency of 113.5\,GHz and with the CD configuration, the PdB
interferometer will not detect emission more extended than about
30\asec, corresponding to $\sim10$\,kpc at a distance of 66\,Mpc.
We can check if such extended emission exists in NGC\,759 by comparing
our single dish flux with the flux obtained with the interferometer. 
Integrating the emission above 0.8\,\jykms\ for the interferometer data,
we obtain a total flux of $37 \pm 2$\ \jykms.
The single dish CO(1--0) spectrum obtained with the IRAM 30--m telescope has
an integrated flux of 11.8\,\kkms\ (Wiklind et al. 1995). Converting the
single dish CO(1--0) flux to Jy using 4.47\,Jy/K, we get $53 \pm 3$\,Jy.
The interferometer thus recovers $\sim70$\% of the single dish flux.
It is possible that 30\% of the emission seen with the single dish
telescope is uniformly extended over more than 10\,kpc; however,
given the uncertainties associated with calibration (about 30\% for
both single dish and interferometric data) and the fact that the 
emission profiles are very similar in shape (see Fig.\,3), we believe
that it is more likely that almost all molecular gas seen with the single
dish telescope is concentrated within the central 5\asec.

\subsection{The velocity field}

The channel maps in Fig.\,1 show strong CO emission in channels at
approximately $\pm 185$\,\kms\ from the center velocity. These channels
correspond to
the two peaks in the double--horned profile shown in Fig.\,3a. A close
inspection of the channel maps shows that the two main emission regions
are separated by a few arcseconds. The separation is, however, of the
order of the angular resolution.
We therefore made Gaussian fits to the individual channels, finding a
systematic shift in center velocity with position, with redshifted emission
coming from the North--East and blueshifted from the South--West (Fig.\,4).
From a fit to the velocities in Fig.\,4 we derive a position angle of
$\sim50^{\circ}$, measured from North to East. 

A position--velocity diagram along a position angle of 50$^{\circ}$ is
shown in Fig.\,5, where the offsets in position are relative to the peak
of the integrated emission. The two peaks of CO emission seen in Fig.\,5
are symmetrical with respect to the center, both in position and velocity.
The separation of the two peaks is 4\asec, corresponding to 1.3\,kpc.
Although the extent is uncertain due to possible residual phase errors,
which tend to enlarge the emission region, we will use 2\ffas0 (650\,pc)
as an `effective radius' of the molecular gas distribution.
This agrees with the average radius of a model fitted to the data
(see Sect.\,5.3).

\subsection{The CO J$=$2--1/J$=$1--0 line intensity ratio}

The extent of the CO emission region of about 4--5\asec\ enables
us to derive a size--corrected CO \jtwo/\jone\ line intensity ratio.
The single dish CO \jone\ and \jtwo\ spectra have
integrated line intensities of 11.8\,\kkms\ and 15.2\,\kkms, respectively.
Convolving the Gaussian telescope HPBW of 23\asec\ for the CO(1--0) line
and 12\asec\ for the CO(2--1) line with a Gaussian shaped source
distribution of HPBW 4\ffas0, we get a \jtwo/\jone\ line ratio of $0.4$.
This value is much lower than what is expected from optically thick and
thermalized CO emission and, since the source size is small compared to
both telescope beams, represents a good estimate.

\subsection{Comparison with radio continuum}

The 1.4\,GHz radio continuum emission of NGC\,759 has been resolved with the
Very Large Array (Feretti \& Giovannini 1994). Its structure is characterized
by diffuse emission without a compact core. The ratio between radio and FIR
fluxes is in agreement with that of star forming galaxies.
A comparison of the distribution of the radio continuum emission and the
molecular gas shows that the continuum emission is centered on the peak
of the molecular gas distribution and that it is contained
within the CO emission region (see Fig.\,1 in Feretti \& Giovannini 1994).

If the continuum emission at 1.4\,GHz were due to free--free emission
and the frequency turn--over lower than 1.4\,GHz, we would expect a flux
density of $\sim10$\,mJy at 113.5\,GHz due to thermal emission in
\hII\ regions. A higher turn--over frequency would yield a higher flux
density. We did not detect any continuum emission at 113.5\,GHz with a 
3$\sigma$ limit of 3.9\,mJy/beam. It is therefore likely that the
continuum emission at 1.4\,GHz is dominated by synchrotron radiation.

\section{Dynamics and kinematics of NGC\,759}

\subsection{A mass model}

In order to interpret our CO data, we need to construct a reasonable model
of the mass distribution in NGC\,759. Since it is classified as an E0 galaxy,
a spherical model with a light distribution following
the de Vaucouleurs $r^{1/4}$ law is a reasonable choice.
We will use the potential--density pair presented by Hernquist (1990), which
closely approximates the de Vaucouleurs $r^{1/4}$ law for elliptical galaxies.
This particular model has the advantage of expressing most aspects of the
dynamics in an analytical form.

The mass as a function of radius is given by
\begin{equation}
M(r) = \frac{M r^{2}}{(r+a)^{2}},
\end{equation}
where $a$ is a parameter related to the effective radius $r_{\rm e}$ within which
half of the luminosity is confined, $r_{\rm e} \approx 1.8153 a$. $M$ is the
total mass, which we derive by first calculating the global $M/L$ ratio.
Following Binney (1982), we have for a spherical $r^{1/4}$ galaxy,
\begin{equation}
\left<\frac{M}{L_{\rm B}}\right> = \frac{0.047\sigma^{2}}{I_{\rm e}R_{\rm e}},
\end{equation}
where $R_{\rm e}$ is the effective radius in kpc , $I_{\rm e}$ is the
surface brightness at $R_{\rm e}$ in units of $\lo\,{\rm pc}^{-2}$
and $\sigma$ is the stellar velocity dispersion in \kms.
The corresponding values for NGC\,759 are $R_{\rm e}=2.6$\,kpc
and $I_{\rm e}=225\,\lo\,{\rm pc}^{-2}$ (Sandage \& Perelmuter 1991).
The velocity dispersion was estimated by Vigroux et al. (1989) to be
$\sigma=250$\,\kms.
Since their value could be influenced by a higher than average velocity
dispersion in the center of NGC\,759, we will use $\sigma=225$\,\kms as
a mean value.
This gives $M/\lb = 4$ in solar units and a mass of
$1.4 \times 10^{11}$\,\mo\ within a radius corresponding to
$R_{25}$ (14.5\,kpc). Extrapolating to $R_{\infty}$, the total mass becomes
$1.7 \times 10^{11}$\,\mo.
The circular velocity as a function of radius is
\begin{equation}
V_{\rm rot} = \frac{\sqrt{GMr}}{r+a}.
\end{equation}
The rotation curve for NGC\,759, using Eq.\,(3), is shown in Fig.\,6.
The maximum rotational velocity of 360\,\kms\ is reached at a
galactocentric radius of $1.4$\,kpc. The rotational velocity at $r=650$\,pc,
the assumed extent of the molecular gas,
is 330\,\kms. Assuming that the molecular gas is distributed in a disk or
ring--structure (see Sect.\,5.3) we derive an inclination of $40^{\circ}$
from the width of the observed emission profile.

\subsection{The gas mass fraction}

The total mass enclosed within a radius of 650\,pc (\mdyn) is given by
Eq.\,(1) as $1.7 \times 10^{10}$\,\mo. This constitutes 10\% of the total
galactic mass and results from the assumption of a constant $M/L$ ratio
and the strongly peaked luminosity distribution of elliptical galaxies.
For elliptical galaxies in general, the $M/L$ ratio tends indeed to be
constant within $\sim 1.2\,R_{\rm e}$ and near a value of 4\,$\mo/\lo$
(Bertola et al. 1993).
The \mhtwo/\mdyn\ ratio becomes 18\% within the central 650\,pc,
where we have included helium at a mass fraction of 1.3.
As long as the mass model for NGC\,759 gives the correct dynamical mass,
the ratio \mhtwo/\mdyn\ only depends on the radius of the molecular gas
distribution.
Whereas the gas mass fraction in the center of NGC\,759 is comparable within
a factor two to the gas mass fractions found in normal spiral galaxies, it
falls to less than 2\% when including the entire galaxy.
The mass of atomic hydrogen is $< 7.4 \times 10^{8}$\,\mo\ (Huchtmeier et al.
1995) and cannot greatly influence the $M_{\rm gas}/M_{\rm dyn}$ ratio.
Hence, NGC\,759 is a gas--poor galaxy when viewed on a global scale.

The \htwo\ mass of NGC\,759 is derived using a standard conversion factor
between integrated CO emission and the column density of \htwo\ (cf. Wiklind
et al. 1995). This ratio is unlikely to be universal, depending on
excitation conditions and metallicity.
For instance, if the metallicity of the molecular gas in NGC\,759 is
low compared to our Galaxy, we may underestimate the mass by a factor of
$\sim$4. This uncertainty in the molecular gas mass does not greatly
influence the results discussed in this paper. The same uncertainty
also applies to other galaxies with which we compare NGC\,759.

\subsection{The CO line profile and the molecular gas distribution}

The double-horned CO profile of NGC\,759 is already in
itself a clue to the molecular gas distribution, ruling out too
large gas concentrations in the very center. It can be shown
by simple integration of the emission along the line of sight
that the bulk of the CO gas should lay outside the rigid part
of the rotation curve to produce the double-horned profile.
For an assumed axisymmetric gas distribution of surface density
$n(r)$, a typical global spectrum $dN/dv (V)$ is the sum over all
radii $r$ of 
\begin{equation}
\frac{dN}{dv} dV = \int{n(r) r dr d\theta} = 
\int{n(r) r dr {{dV}\over{V_{\rm rot}(r) \sin{\theta} \sin{i}}}},
\end{equation}
where $V_{\rm rot}(r)$ is the rotational velocity at radius
$r$, and $i$ the inclination of the galaxy in the sky ($i=0$
is face--on). Since the observed $V = V_{\rm rot} \cos{\theta} \sin{i}$,
it is straightforward to derive that in the simple but
unphysical case of a constant rotational velocity $V_{\rm rot}$
at all radii, the double--horn profile
\begin{equation}
\frac{dN}{dv} \propto \left[1 - \left(\frac{V}{V_{\rm rot}
\sin{i}}\right)^{2}\right]^{-1/2},
\end{equation}
results whatever the radial distribution of the surface
density $n(r)$. In the opposite case of a rigid body rotation
curve, and with a constant gas  surface density $n(r)=n_0$
until $R_{\rm max}$, where the rotation curve turns over to
become flat, the spectrum becomes
\begin{equation}
\frac{dN}{dv} \propto \left[1 - \left(\frac{V}{V_{\rm max}
\sin{i}}\right)^2 \right]^{1/2},
\end{equation}
where $V_{\rm max}$ is the rotational velocity reached at
$R_{\rm max}$.
These cases represents two extremes of rotation curves and
show that a parabolic profile (large $R_{\max}$) will
transform into a double--horned when $R_{\rm max}$ decreases.
The depth of the double-horn profile is thus a good indicator
of the gas concentration with respect to the turn--over of the 
rotation curve.
Emission profiles obtained from Eqns. (5) and (6) are shown in Fig.\,7.

\smallskip 

We have modeled in more detail the gas distribution and 
kinematics to better determine the constraints brought by the
CO observations.
We have assumed that the gas is in cylindrical rotation in a
thick disk or ring, in the potential of the elliptical galaxy,
as fixed  in Sect.\,5.1, also accounting for the potential of
the gas itself. The latter is not a negligible
contribution, given its high central condensation.
For the disk we choose an analytical density--potential pair in the
form of Toomre disks of order n$=$2 (cf Toomre 1964). 
To simulate a disk, we choose the surface density:
\begin{equation}
n(r) = n_0 \left(1 + {{r^2}\over{d^2}}\right)^{-5/2},
\end{equation}
where {\em d} is the disk characteristic scale, and the corresponding 
rotational velocity is
\begin{equation}
V^2(r) = {{8 \pi G n_0}\over{3 d}} r^2 \left(1 + {{r^2}\over{4 d^2}}\right)
\left(1 + {{r^2}\over{d^2}}\right)^{-5/2}.
\end{equation}
For a ring, we choose the difference of two such distributions
with $d_1$ and $d_2$ as characteristic scales, and with the
same central surface density $n_0$.
Since the total gas mass is fixed ($\mhtwo = 2.4 \times 10^{9}$\,\mo),
and so is the elliptical potential, we only vary the scales 
$d_1$ and $d_2$. The global spectrum is then derived by a simple 
radial integration.
A best fit with the observed spectrum is obtained for a ring
constructed with scale lenghts $d_1=800$\,pc and $d_2=750$\,pc,
viewed at an inclination $i = 40^{\circ}$ (see Fig.\,8). This
gives a maximum surface density of 750\,\mopcsq\ at $r=500$\,pc.
The surface density profile is shown in Fig.\,9.
Although the ring has a maximum at 500\,pc, the distribution
is extended at galactocentric distances up to $\sim$1.5\,kpc
and the radius of 650\,pc inferred from the interferometer
data represents an average radius for the gas distribution.

Is this extended gas distribution in agreement with the 
barely resolved interferometric map? We have answered
the question by simulating the ring with $N=$10 000 particles,
distributed as above,  with a cylindrical rotation in a
disk thickened by a ${\rm sech}^2 (z/z_0)$ law, with
$z_0 = 200$\,pc, and with an isotropic velocity dispersion
of 30\,\kms. We then `observed' the distribution with an
inclination $i =  40^{\circ}$, and an elliptical beam of
3\ffas1$\times$2\ffas3. The total integrated density of
the model distribution is shown in Fig.\,10, and is in
good agreement with the observations.

%--------------------------Table 3--------------------------------------------
\begin{table*}
\begin{flushleft}
\caption[]{\htwo\ densities and \tb(1--0) for different fractional CO abundances
and for CO \jtwo/\jone\ = 0.4}
\scriptsize
\begin{tabular}{|c|ccc|ccc|ccc|}
\hline
\multicolumn{1}{|c}{ }                    &
\multicolumn{8}{c}{ }                     &
\multicolumn{1}{c|}{ }                    \\
\multicolumn{1}{|c}{ }                    &
\multicolumn{9}{c|}{$X$(CO)/(d$V$/d$r$)\ $(\kmspc)^{-1}$}\\
\multicolumn{1}{|c}{ }                    &
\multicolumn{8}{c}{ }                     &
\multicolumn{1}{c|}{ }                    \\
\multicolumn{1}{|c}{ }                    &
\multicolumn{3}{c}{$10^{-6}$}             &
\multicolumn{3}{c}{$10^{-5}$}             &
\multicolumn{3}{c|}{$10^{-4}$}            \\
\hline
 & & & & & & & & & \\
\multicolumn{1}{|c|}{\tkin}               &
\multicolumn{1}{c}{\tb(1--0)$^{a)}$}      &
\multicolumn{1}{c}{\nhtwo}                &
\multicolumn{1}{c|}{\mhtwo/\lco}          &
\multicolumn{1}{c}{\tb(1--0)$^{a)}$}      &
\multicolumn{1}{c}{\nhtwo}                &
\multicolumn{1}{c|}{\mhtwo/\lco}          &
\multicolumn{1}{c}{\tb(1--0)$^{a)}$}      &
\multicolumn{1}{c}{\nhtwo}                &
\multicolumn{1}{c|}{\mhtwo/\lco$^{a)}$}   \\
\multicolumn{1}{|c|}{ }                   &
\multicolumn{2}{c}{ }                     &
\multicolumn{1}{c|}{ }                    &
\multicolumn{2}{c}{ }                     &
\multicolumn{1}{c|}{ }                    &
\multicolumn{2}{c}{ }                     &
\multicolumn{1}{c|}{ }                    \\
\multicolumn{1}{|c|}{K}                   &
\multicolumn{1}{c}{K}                     &
\multicolumn{1}{c}{\cmcb}                 &
\multicolumn{1}{c|}{\mo/\kkms}            &
\multicolumn{1}{c}{K}                     &
\multicolumn{1}{c}{\cmcb}                 &
\multicolumn{1}{c|}{\mo/\kkms}            &
\multicolumn{1}{c}{K}                     &
\multicolumn{1}{c}{\cmcb}                 &
\multicolumn{1}{c|}{\mo/\kkms}            \\
\hline
 & & & & & & & & & \\
10 & 2.8 & 1144 & 25 & 5.3  & 464 &  9  & 4.2 & 180    & 7 \\
20 & 1.3 &  422 & 33 & 3.2  & 216 & 10  & 4.0 & $<$100 & $<$5 \\
30 & 0.9 &  276 & 39 & 2.6  & 153 & 10  & --- & ---    & ---  \\
40 & 0.7 &  210 & 43 & 2.1  & 121 & 11  & --- & ---    & ---  \\
60 & 0.6 &  152 & 52 & ---  & --- & --- & --- & ---    & ---  \\
 & & & & & & & & & \\
\hline
\end{tabular}
\\
a)\ \tb\ is the excess brightness temperature above the microwave background. \\
b) \ $\mhtwo/\lco=2.1\sqrt{\nhtwo}/\tb(1-0)$\ \ (cf. Radford et al. 1991).
\end{flushleft}
\end{table*}
%----------------------End Table 3-------------------------------------------- 

%9
\begin{figure}
\psfig{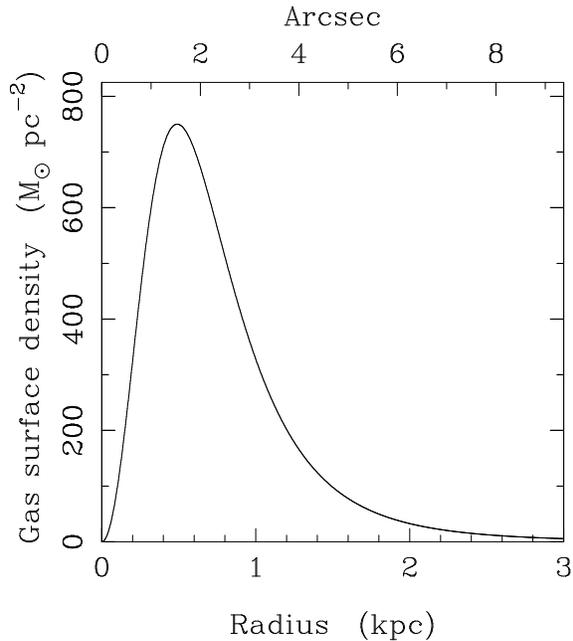}
%\picplace{6.0cm}
\caption[]{The gas surface density as a function of galactocentric radius,
obtained using the ring constructed from two Toomre disks (cf. Sect.\,5.3).}
\end{figure}

%10
\begin{figure}
\psfig{figure=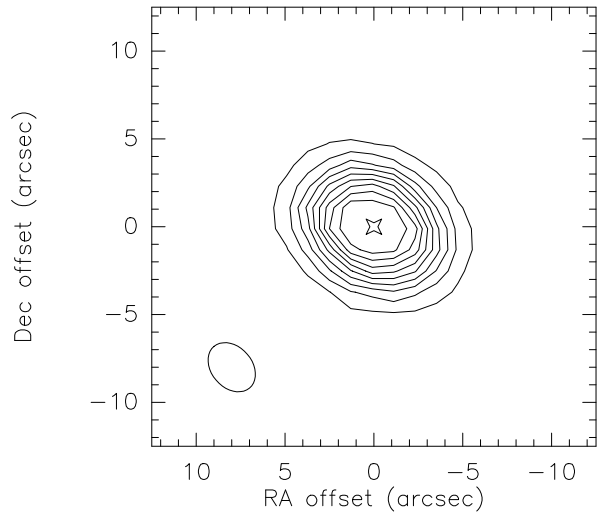,bbllx=65mm,bblly=35mm,bburx=175mm,bbury=170mm,width=8.5cm,angle=-90}
%\picplace{5.0cm}
\caption[]{A synthetic image of the ring--structure seen at
an inclination of 40$^{\circ}$, convolved with a Gaussian beam of the
same HPBW size as the CLEANed beam ($3\ffas1\times2\ffas3$).}
\end{figure}

%11
\begin{figure}
\psfig{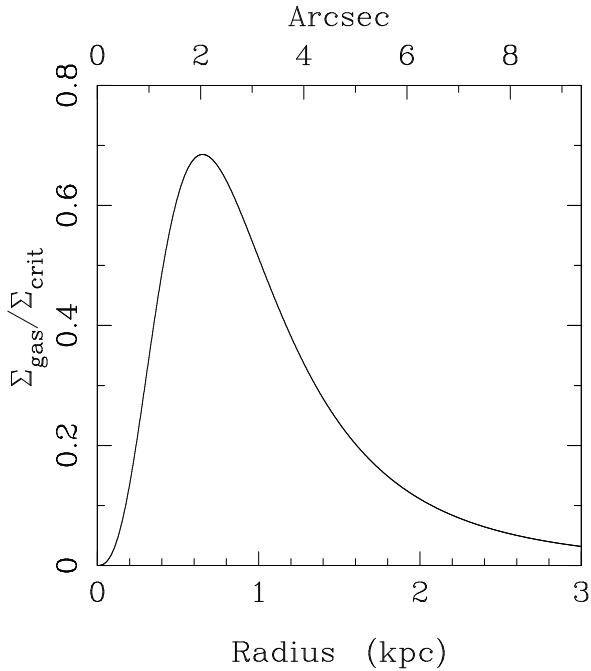}
%\picplace{5.0cm}
\caption[]{The ratio of the molecular gas surface density $\Sigma_{\rm gas}$
(including helium at a mass fraction of 1.3) and the critical gas surface
density $\Sigma_{\rm crit}$, as a function of galactocentric radius for a
galaxy described by the mass model presented in Sect.\,5.1.}
\end{figure}

\section{Discussion}

\subsection{Is NGC\,759 a starburst elliptical?}

Apart from the mere presence of molecular gas, there are
three facts which suggest that massive star formation
is taking place in the inner 650\,pc of NGC\,759:
{\em (i)} optical emission lines from \hII\ regions, {\em (ii)}
diffuse radio continuum emission and {\em (iii)} FIR emission
which most likely comes from dust heated by massive stars.
With a Salpeter initial mass function, the inferred star
formation rate (SFR) is $\sim$7\,\sfrat\ (Table\,1).

How does the molecular data fit into the picture of
NGC\,759 as an actively star forming elliptical galaxy?
Below we discuss the molecular properties in more detail
in view of its implications for the star formation activity
in NGC\,759.

\subsection{The molecular gas surface density and $\Sigma_{\rm crit}$}

Kennicutt (1989) has argued for the existence of massive star formation
activity in spiral galaxies only when the gas surface density exceeds a
critical value, for which the gaseous disk is expected to become unstable
to large scale density perturbations. Using the Toomre (1964) formulation,
the critical surface density can be expressed as:
\begin{equation}
\Sigma_{\rm crit} =  \alpha \frac{\kappa c}{\pi G},
\end{equation}
where $\alpha$ is a constant of order unity (0.7 for spiral galaxies),
$\kappa$ is the epicyclic frequency, $c$ is the gas velocity dispersion and
$G$ the gravitational constant.
Using the rotation curve corresponding to the mass model presented in
Sect.\,5.1, the critical gas surface density becomes
\begin{equation}
\Sigma_{\rm crit} \approx 1.6 \frac{V_{\rm rot}}{r}
\sqrt{3 - \frac{2 r}{r+0.55r_{\rm e}}}\ \mopcsq,
\end{equation}
where $r$ is the galactocentric distance expressed in kpc, $V_{\rm rot}$
is the rotational velocity in \kms, and we have assumed a gas velocity
dispersion of 30\,\kms. The ratio of the molecular gas surface density
obtained from the model distribution presented in Sect.\,5.3 and the
$\Sigma_{\rm crit}$ derived using Eq.\,(10) is shown in Fig.\,11.
Due to the rapid increase of $\Sigma_{\rm crit}$ with decreasing radius, the
ratio is everywhere less than one. Despite the high molecular gas surface
density we do not expect the gas to be highly susceptible to large scale
density perturbations. The ratio is, however, of order unity around the peak
of the molecular ring, and this allows gravitational instabilities to
develop on the same scale as in the disk of normal spiral galaxies.
The maximum of $\Sigma_{\rm gas}/\Sigma_{\rm crit}$ falls at a slightly
larger galactocentric distance than the peak of the molecular ring.
A thin ring of \hII\ regions should therefore be visible in narrow band
\halfa\ images at a distance of $\sim$2\ffas0 from the nucleus of NGC\,759.

\subsection{$\Sigma_{\rm H_{2}}$ and the \lfir/\mhtwo\ ratio}

The molecular gas surface density of 750\,\mopcsq\ found for
NGC\,759 is high when compared with the Galactic molecular ring
(12\,\mopcsq) and the Galactic center region ($200-300$\,\mopcsq),
but it is a factor $2-3$ smaller than the surface density obtained
by assuming a constant surface density disk truncated at a radius
650\,pc. In the latter case NGC\,759 would have a gas surface
density similar to those of ultraluminous FIR galaxies
(cf. Scoville et al. 1991).
The difference comes from the larger extension of the CO
distribution at low surface density (out to 1.5\,kpc) in the
modelled ring distribution (see Sect.\,5.3).

In Fig.\,12 we compare the \lfir/\mhtwo\ ratio vs.
$\Sigma_{\rm H_{2}}$ for ultraluminous FIR galaxies and NGC\,759,
with $\Sigma_{\rm H_{2}}$ derived assuming a truncated
gas disk and using the virial theorem.
NGC\,759 has an \lfir/\mhtwo\ ratio of 4.7\ \lo/\mo, which is
much smaller than for ultraluminous FIR galaxies who typically
have ratios in excess of 40\ \lo/\mo.
In fact, the \lfir/\mhtwo\ ratio of NGC\,759 is only half the
average value for normal IRAS selected spiral galaxies (cf.
Young et al. 1989).
The large $\Sigma_{\rm H_{2}}$ of NGC\,759 and the low
\lfir/\mhtwo\ ratio suggests that NGC\,759 is underluminous
in \lfir\ with respect to its \htwo\ mass as compared to
ultraluminous FIR galaxies. This could be caused by a lack
of massive stars heating the dust in NGC\,759, suggesting a
low star formation efficiency, or the presence of an extra
heating source of the dust in ultraluminous FIR galaxies,
possibly an AGN. 
If the metallicity of the molecular gas in
NGC\,759 should be very low, which it could be if the gas was accreted
from a dwarf galaxy, the \htwo\ mass would be underestimated, enhancing
the difference between NGC\,759 and the ultraluminous FIR galaxies.

\subsection{Subthermal excitation of the CO emission}

A CO \jtwo/\jone\ line ratio as low as 0.4 (Sect.\,4.3) is rarely
encountered in molecular clouds, neither in our Galaxy nor in
external spiral galaxies (Braine \& Combes 1992). The low ratio
implies a subthermal excitation of the J$=2$ level and is usually
interpreted as cold and/or diffuse molecular gas.

Table\,3 summarizes large velocity gradient calculations for a CO
\jtwo/\jone\ ratio of 0.4. The calculations are taken from Castets
et al. (1990) and give for three different fractional CO abundances
per velocity interval, $X({\rm CO})/(dV/dr)$\,$(\kmspc)^{-1}$, the
brightness temperature \tb, the \htwo\ density and the ratio
\mhtwo/\lco. $X({\rm CO})$ is equal to $n({\rm CO})/n({\rm H_{2}})$.
A small fractional CO abundance can be interpreted as a low metallicity
gas and a large abundance as a high metallicity gas. 
The \mhtwo/\lco\ ratio is derived in the same way as in
Radford et al. (1991). A value of 4.5\,\mo/\lo\ corresponds to
the standard $N_{\rm H_{2}}/I_{\rm CO}$ conversion ratio.

A fractional CO abundance lower than $10^{-6}$\,$(\kmspc)^{-1}$ can be
excluded on grounds of a very low brightness temperature and an
\mhtwo/\lco\ ratio $\sim$10$^3$\,\mo/\lo.
In the case of a small fractional CO abundance we underestimate the
molecular mass by a factor 5--10, whereas a high fractional abundance
gives the correct mass to within a factor of 2.
In the simple hypothesis of a homogeneous one--component medium,
under the LVG approximation, two extreme physical conditions are
possible for the molecular gas: either cold and dense
($\tkin \approx 10$\,K, $\nhtwo \ga 10^3$\,\cmcb)
or warm and diffuse ($\tkin \approx 40$\,K,
$\nhtwo \approx 10^2$\,\cmcb).
A more realistic solution might be a multi--medium composed of
dense clumps embedded in a diffuse component, where the CO
emission is dominated by the diffuse component
while most of the mass is contained in the dense clumps.
This can be seen from the higher \mhtwo/\lfir\ ratio of the
cold and dense component compared to the diffuse gas (see Table\,3).

\subsection{Comparing with a bonafide merger: NGC\,7252}

A good example of an advanced merger is NGC\,7252, which shows evidence of
a recent starburst in its center (cf. Schweizer 1982). Although the merger
initially consisted of two disk galaxies, the overall light distribution now
follows the $r^{1/4}$ law (Schweizer 1982, Stanford \& Bushouse 1991).

Dupraz et al. (1990) observed NGC\,7252 in the CO \jone\ and \jtwo\ lines
with  a single dish telescope (SEST), and Wang et al. (1992) observed it
with the Owens Valley interferometer, achieving a resolution of
3\ffas8$\times$2\ffas1 ($1.1 \times 0.6$\,kpc, assuming a distance of
63\,Mpc).
The molecular gas distribution in NGC\,7252 has a striking resemblance to
that of NGC\,759.
The CO profiles from the single dish and the interferometer are double--horned,
and the inclination of the gas appears to be around 40$^{\circ}$.
At least 72\% of the CO(1--0) emission in NGC\,7252 comes from within a
radius of 1.5\,kpc. The CO \jtwo/\jone\ ratio, corrected for the size
of the emission region, is $0.4-0.6$ (depending on how much of the CO(2--1)
emission that is extended). The molecular gas surface density of NGC\,7252
is similar to NGC\,759, when calculated in the simple way discussed in
Sect.\,5.2.
However, while the dust temperatures are comparable, the \lfir/\mhtwo\ ratio
of NGC\,7252 is a factor 3 higher than in NGC\,759

Since the projected light distribution of NGC\,7252 closely follows the
$r^{1/4}$ law, it is appropriate to use a mass model similar to the one
we used for NGC\,759 (Sect.\,5.1). From the light profile presented by
Stanford \& Bushouse (1991) we derive an effective radius $r_{\rm e}$ of
10\ffas8 (3.3\,kpc). If we express the total mass of NGC\,7252 as a function
of the mass--to--light ratio,
\begin{equation}
M_{\rm tot} = \left<\frac{M}{L_{\rm B}}\right> \lb,
\end{equation}
and use Eq.\,(1) and (2), we get the following expressions for the mass
within radius $r=1.5$\,kpc and the corresponding rotational velocity,
\begin{equation}
M(r) = 9.4 \times 10^{9} \left<\frac{M}{L_{\rm B}}\right>\ \mo,
\end{equation}
\begin{equation}
V(r) = 166 \sqrt{\left<\frac{M}{L_{\rm B}}\right>}\ \kms.
\end{equation}
The \htwo\ mass within a radius of 1.5\,kpc is $3 \times 10^{9}$\,\mo\ (Wang
et al. 1992), giving a \mhtwo/\mdyn\ ratio of 
$0.3 \left<M/L_{\rm B}\right>^{-1}$.
The mass--to--light ratio must be at least 3 to give the observed velocity width
($i=90^{\circ}$). This gives a molecular gas mass fraction of $\la 10$\% in the
inner region of NGC\,7252. This is considerably smaller than the 45\% derived
by Wang et al. (1992). 
If the inclination is in the range $40^{\circ}-50^{\circ}$,
as implied by the interferometer map, the mass--to--light ratio is 
$5-7$, further diminishing the molecular gas mass fraction.

Apart from the \lfir/\mhtwo\ ratio, the advanced merger NGC\,7252 and the
elliptical NGC\,759 show striking similarities in their molecular gas
properties.

The very high gas mass fractions found in the central regions of
ultraluminous FIR galaxies (cf. Scoville et al. 1991), including
NGC\,7252, have been obtained from a dynamical mass derived using
the virial theorem. This is applicable to disk systems with a small
central mass concentration compared to elliptical galaxies.
However, most of these ultraluminous FIR galaxies are mergers, and
presumably have a light distribution following the $r^{1/4}$ law.
Hence, it would be appropriate to use a more concentrated mass model
such as that described by Hernquist (1990). This would decrease the
central gas mass fractions in ultraluminous FIR galaxies by a factor
$3-10$.

%12
\begin{figure}
\psfig{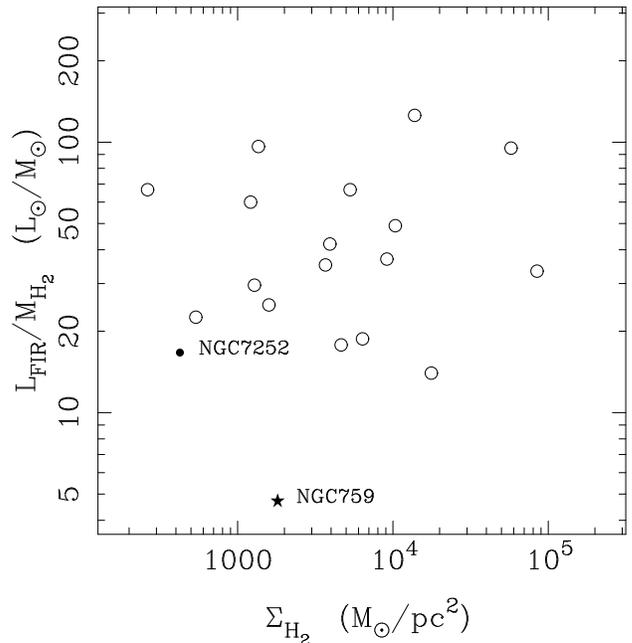}
%\picplace{6.0cm}
\caption[]{The \lfir/\mhtwo\ ratio as a function of the average nuclear
molecular gas surface density for ultraluminous far--infrared galaxies
(open circles), NGC\,7252 (filled circle) and NGC\,759 (star).
The gas surface density has been derived in the same manner for all
galaxies; using a truncated gas disk and the virial theorem.}
\end{figure}

\section{Summary and conclusions}

NGC\,759 is an elliptical galaxy which contains $2.4 \times 10^{9}$\,\mo\ of
molecular gas, most of it confined to a circumnuclear ring with an average
radius of $\sim$650\,pc.
The average molecular gas surface density in the center region of NGC\,759 is
750\,\mopcsq. Although this value is high, it is always less 
than the critical gas surface density for large scale gravitational 
instabilities.
The low CO \jtwo/\jone\ line ratio of 0.4, is consistent with a
molecular gas consisting of cold and dense clumps
($\tkin \approx 10$\,K, $\nhtwo \ga 10^3$\,\cmcb) embedded in a warm and
diffuse molecular gas ($\tkin \approx 40$\,K, $\nhtwo \approx 10^2$\,\cmcb).
The dense component contains most of the molecular gas mass, while
the diffuse gas dominates the emission.
Finally, when compared with galaxies of similar gas surface densities, NGC\,759
appears to be underluminous in \lfir\ with respect to its molecular gas mass by
a factor $3-10$.

\medskip

In a quiet environment, molecular gas condenses into a very clumpy structure
with local high densities and a low volume filling factor, resulting in
a low average gas surface density. Intense star formation partially destroys
the clumpy structure and the fraction of diffuse molecular gas is considerably
enhanced. Consequently the filling factor and the average surface density will
also increase. Although the kinetic temperature increases due to these 
processes, the CO excitation temperature can become quite low due to 
subthermal excitation caused by the low volume density of \htwo.

This scenario, which seems to be applicable to NGC\,759, implies that star
formation is ongoing and has been so long enough to influence the molecular
gas properties in the circumnuclear ring.
The low \lfir/\mhtwo\ ratio could have its explanation in the lack of an
additional heating source for the dust in the form of an AGN.

\medskip

The high concentration of molecular gas in the center of NGC\,759, the
lack of atomic gas and the relatively modest amount of star forming activity
is reminiscent of the conditions found in the merger NGC\,7252.
The main difference being that while NGC\,7252 exhibits gas--rich tidal
tails and a high star formation efficiency, NGC\,759 shows no sign of 
gravitationally induced disturbances and has a low star formation efficiency.

The cluster A262 contains a high fraction of spiral galaxies (75\%).
The spirals are less centrally concentrated and have a higher line--of--sight
velocity dispersion then the ellipticals (Moss \& Dickens 1977).
NGC\,759 is situated in outer parts of A262, at a projected distance of
1.4\,Mpc from the cluster center.
Could NGC\,759 represent a very late stage in the merging of two disk galaxies?
The fate of the gaseous component in the merging of two gas rich galaxies is
two--fold: a major part of the gas looses angular momentum quickly and falls
towards the center while the merging process is on--going. If this gas
component is capable of forming stars, it can produce the high stellar
phase--space density characteristic for elliptical galaxies. The other
part of the gas is ejected into tidal arms, which remain bound to the
system (Hernquist \& Spergel 1992; Hibbard \& Mihos 1995). This gas
will fall back to the galaxy at later times.
If elliptical galaxies are formed through the merging of two gas rich disk
galaxies, an important question is: where are the galaxies which are in between
the final product, an inert and gas--free elliptical galaxy, and those which
still show signs of the merging process, such as NGC\,7252. This question was
first raised by I. King and has been named the `King Gap'.
Galaxies in the `King Gap' should look like proverbial elliptical but still 
retain some signatures of the merging process. These signatures can be:

\begin{itemize}

\item Gas and dust components as well as an intermediate aged stellar population.

\item Optical fine structure in the form of faint shells, ripples and boxy isophotes
(cf. Schweizer \& Seitzer 1992).

\item Kinematically distinct sub--systems (cf. Surma \& Bender 1995).

\item Gas in the outer part of the tidal arms require very long time scales
to return to the galaxy. Hence, one would expect the presence of low HI
column density at large galactocentric distances (cf. Hibbard \& Mihos 1995).

\end{itemize}

NGC\,759 does show unusual gas properties for an elliptical galaxy and if it
represents a galaxy in the `King Gap', one or more of the additional
signatures should be found in a more extended study.

\acknowledgements
We acknowledge the efficient support from the staff at Plateau de Bure
and IRAM Grenoble during the course of these observations.

%_____________________________________________________________________

\end{document}